\documentclass[prl,aps,twocolumn,showpacs]{revtex4}
\usepackage{graphicx}
\usepackage{amsmath}
\newcommand \beq  {\begin{equation}}
\newcommand \eeq  {\end{equation}}
\newcommand \bea {\begin{eqnarray} }
\newcommand \eea {\end{eqnarray}}

\begin{document}
\title{LiV$_2$O$_4$: frustration induced heavy fermion metal}
\author{ J. Hopkinson and P. Coleman}
\address{Center for Materials Theory,
Department of Physics and Astronomy, 
Rutgers University, Piscataway, NJ 08854, USA.
}

\begin{abstract}
We propose a two-stage spin-quenching scenario for the unusual heavy
fermion state realized in the mixed valent metal LiV$_2$O$_4$.  In
this theory, local valence fluctuations are responsible for the
formation of partially quenched, spin $\frac{1}{2}$ moments below room
temperature{\cite{kondo}}. Frustration of the intersite spin couplings
then drives the system to produce the heavy Fermi liquid seen at low
temperatures.{\cite{takagi}} The anomalous resistivity and the sign
change of the Hall constant can be understood naturally within this
model, which also predicts a unique symmetry for the heavy
quasiparticle bands that that may be observed in de Haas-van Alphen
experiments.
\end{abstract} 


\maketitle
Magnetic frustration has attracted increasing 
attention in recent years as a new 
tool to suppress antiferromagnetism and generate new types
of electronic behavior.  
One exciting possibility is the development of frustration induced
heavy electron behavior 
in d-electron systems where the Kondo temperature is generally 
too small to overcome magnetic order 
without frustration.
The recognition of heavy
fermion behavior in LiV$_2$O$_4${\cite{kondoprl}} was an important
first step in this direction.  
In
LiV$_2$O$_4$, the frustrated lattice remains  undistorted to the lowest temperatures
measured,{\cite{xray}} 
The heavy electron state which then develops 
is unusual in many respects: it displays 
a monotonically increasing resistivity{\cite{takagi}}, a
pressure-driven metal-insulator transition{\cite{takag}} and field
independent heat capacity up to 30T{\cite{takag}}.



LiV$_2$O$_4$ forms a spinel structure.  The magnetic vanadium atoms
are homogeneously mixed valent forming  a lattice of undistorted
corner-shared tetrahedra. The spin at each
site fluctuates between S = 1 (3d$^2$) and S = $\frac{1}{2}$ (3$d^1$)
with a formal valence of $3.5$.
Antiferromagnetic couplings between these spins{\cite{muhtar,lee}} imply
that the lattice is a highly frustrated metal.  There is a strong
on-site Hund's coupling between spins of order 1 eV{\cite{anisimov}},
which is manifested in the photoemission{\cite{fujimori}} and the high
temperature magnetic susceptibility{\cite{muhtar}}.  At room
temperature the magnetic suceptibility and specific heat indicate the
presence of spin  $\frac{1}{2}$ local moments at every
site.{\cite{kondo,takagi}}. Below 4K, the moments quench into a heavy
Fermi liquid.{\cite{kondo}}

There have been several attempts to explain this physics.  One class
of model sees the frustration as secondary,
the essential physics being due to mixed valence{\cite{varma}},
exhaustion{\cite{anisimov}}, 1D chains{\cite{fujimoto}} or a two-band
Hubbard model{\cite{kusunose}}.  Another argues that frustration is
essential, leading to a tetrahedron
rule{\cite{fulde}}, a frustrated Kondo lattice{\cite{burdin}},
antiferromagnetically coupled spins Hund's coupled to conduction
electrons{\cite{lacroix}} or quantum criticality{\cite{shannon}}.
Here we present a simple two-band model which we believe captures the
essential physics: a two stage spin quenching that could not occur in
the absence of frustration; and show how our results can be extended
to the pyrochlore structure.

The high temperature physics of LiV$_2$O$_4$ can be understood as an
S=1 Mott insulator doped with holes on a pyrochlore lattice.  The
strong Hund's coupling leads us to consider only spin-1 and
spin-$\frac{1}{2}$ sites coupled antiferromagnetically over the
lattice via direct exchange{\cite{singh}}, 
\bea H &=& -t\sum_{(i,j)}
(X_{0 \sigma}(i) X_{\sigma 0}(j) + h.c.) + \sum_{(i,j)} J s_i\cdot
s_j 
\eea 
where 
$X_{0 \sigma} =
|\frac{1}{2}, \sigma><1,\sigma'| = \chi^{\dagger} b_{\sigma}$ is a
Hubbard operator allowing the charged holes to move
in the spin background via a hopping matrix element $t$, 
subject to the constraint $b_{\sigma}^{\dagger}b_{\sigma} +
\chi^{\dagger}\chi = 2$.  $J$ is an antiferromagentic coupling
between nearest neighbor spins.
The local physics of this model is that of a mixed valent impurity 
undergoing charge fluctuations $d^{2} \rightleftharpoons d^{1}+e^{-} $
via hybridization with the surrounding bath of holes, as might be
formalized in a dynamical mean-field description of the above model.
The effective impurity 
Hamiltonian replacing the first term in $H$ would then be
\begin{equation}\label{}
H_1 = \sum_{k \sigma} \epsilon_k c^{\dagger}_{k \sigma} c_{k \sigma} +
\sum_{k \sigma}t(X_{\sigma 0}c^{\dagger}_{k \sigma} + h.c.) . 
\end{equation}
As the frustration of the pyrochlore lattice
prevents magnetic ordering from occuring, this term dominates the high
temperature physics.  From strong coupling
arguments{\cite{krishna-murthy}}, 
we know that a Fermi sea of
electrons forms below $T_1$, quenching  the local moment character on each site to
S=$\frac{1}{2}$.  This provides a natural explanation for the behavior
seen below room temperature. 
On the lattice,  this fluid can be visualized as a smooth Fermi sea,
from which the residual spin $\frac{1}{2}$ moments protrude 
like spines on the hide of a porcupine. (Fig. 1.)
The original antiferromagnetic
interaction between nearest neighbor sites now couples the electron 
sea to the residual spin $\frac{1}{2}$ moments. 
The properties of this ``porcupine'' Fermi sea are the topic of the
remainder of this paper.  
The question is  -- {\it{what is the
nature of the intrasite screening process that removes the free spins 
(spines) at low temperatures?}}

At low temperatures, $J$ becomes an effective Kondo coupling between
the localized spins and the newly freed electrons in the
Fermi sea.  The locally symmetric spin state required to satisfy the
strong Hund's coupling means that spatial overlap of the realized
Wannier states is forbidden as the heavy quasiparticle forms at low temperatures.  Thus, even
after the complete break-down of any orbital picture, as the mixed
valent spin quenching of the upper level occurs in the totally
symmetric channel the second stage spin quenching can occur only in
an orthogonal channel.  This is the analogue of the atomic Hund's
selection rule for itinerant states.  Thus the conditions
\bea
\begin{cases} \text{no overlap} \\ \nonumber \text{no phase transition (equivalent sites)} \\ \nonumber \text{no time reversal symmetry breaking} 
\end{cases} 
\eea
arise, uniquely specifying the second channel (Fig.2).
The unusual symmetry of the composite quasi-particles resulting from this second spin-quenching should be observable by de Haas - van Alphen experiments.

\begin{figure}
\includegraphics[scale=0.4]{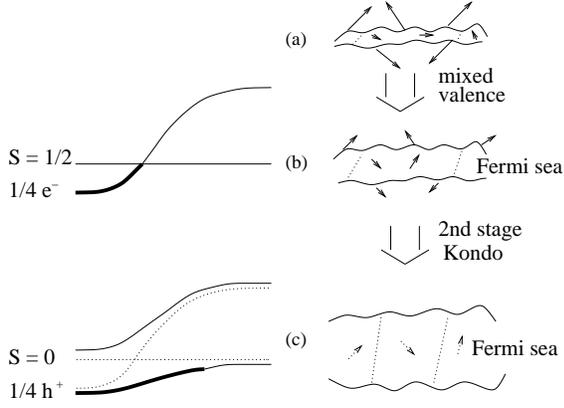}
\caption{\label{figure}  
(a) High temperature phase with large moments. (b)
Intermediate temperature phase
with underscreened $S=1/2$ moments embedded in a Fermi sea
like the spines of a `Porcupine' (left: 
corresponding  $\frac{1}{4}$ filled band interacting with a localized level) (c)
Low temperature phase in which the 
Heisenberg interaction hybridizes these bands leaving a heavy Fermi
liquid of holes. (left: corresponding band picture)}
\vskip -0.5cm
\end{figure}


We now treat the second stage at the mean field level keeping in mind that for LiV$_2$O$_4$, we really want to have a half-filled localized d-band interacting with a quarter-filled delocalized d-band.  Estimates of $J \approx \theta_{CW}^{high T} \approx 400 K${\cite{muhtar}}, T$_{c2} \approx 4-50 K$ and $t \approx 2 eV${\cite{singh}} (band theory) have been made, and we expect that $T_{c1} \propto t$ and $T_{c2} \propto t e^{\frac{\# tb^2}{J}}$ as these are the two energy scales in the problem.
It is convenient to recast the Hubbard operators into a slave boson language, 
\begin{eqnarray}\label{}
\nonumber H &=& \sum_{<ij>}(-tX_{0 \sigma}^iX_{\sigma 0}^j +  J
{\bf{s}}_i \cdot {\bf{s}}_j) \\
&-& J_H \sum_j ({\bf{s}}_j^2 -2+\frac{5}{4}X_{00}^{j}) 
\end{eqnarray}
where $X_{0 \sigma} = b^{\dagger}c_{\sigma}$ is a Hubbard operator
allowing the holes to move in the upper e$_g'$ symmetry channel,
${\bf{s}} = \begin{pmatrix}c_{\alpha}^{\dagger} &
d_{\alpha}^{\dagger}\end{pmatrix}\frac{\sigma_{\alpha
\beta}}{2}\begin{pmatrix}c_{\beta} \cr d_{\beta}\end{pmatrix}$ is the
total spin, and we need to enforce the constraints $n_{c} + n_b = 1$,
$n_{d} = 1$ (a localized spin-$\frac{1}{2}$ level of $a_{1g}$
symmetry). The Hund's coupling $J_{H}$ is to
be taken to infinity to enforce the constraint $S^2 = 2$ at 
doubly occupied sites. 

We now demonstrate how a mean field treatment of Eq.3 leads us to the effective Hamiltonian:
\bea
H_{mf} &=& \sum_{k \sigma}\begin{pmatrix}{\mathcal{C}}^{\dagger}_{k \sigma} &{\mathcal{D}}^{\dagger}_{k \sigma}\end{pmatrix} {\bf{h}}_k\begin{pmatrix}{\mathcal{C}}_{k \sigma} \cr {\mathcal{D}}_{k \sigma} \end{pmatrix}   + 2\sum_{\Gamma}\frac{\bar \Delta_{\Gamma}\Delta_{\Gamma}}{J^{\Gamma}} \\ \nonumber &+& \lambda_1(|b|^2 - 1) -  \lambda_2
\eea
where 
\bea
{\bf{h}}_k = \begin{pmatrix}-2t|b|^2 \Phi^{(1)}_k + \frac{\lambda_1-\mu}{N} &\Delta_{\Gamma}\Phi_k^{* (\Gamma)} + i\xi \cr\bar \Delta_{\Gamma}\Phi_k^{(\Gamma)} + i\bar\xi  &  \frac{\lambda_2}{N}\end{pmatrix}
\eea
and
${\mathcal{D}}^{\dagger}_k = \begin{pmatrix}d_{1k} & .. &d_{Nk}\end{pmatrix},  \hspace{2mm} {\mathcal{C}}^{\dagger}_k = \begin{pmatrix}c_{1k} &.. &c_{Nk}\end{pmatrix}$
demark the number of bands arising from the magnetic unit cell of the
lattice under consideration.  $\lambda_1$ and $\lambda_2$ are Lagrange
multipliers fixing the occupancies of the
electron and spin liquid. $\Phi_k^{(\Gamma)}$ is an N dimensional 
matrix representing the Wannier functions, where N is the number of
atoms per unit cell 
and $\Gamma = 1,..,M$ specifies the number of orthogonal channels
required to decouple the Heisenberg interaction between one site and
its $M$ nearest neighbors.

The mean field kinetic energy term of $H$ is
\bea
H_{kin} = -\sum_{<ij>}t(c^{\dagger}_{\sigma}b)_i(b^{\dagger}c_{\sigma})_j=-2t|b|^2 \sum_k {\mathcal{C}}^{\dagger}_k \Phi^{(1)}_k {\mathcal{C}}_k 
\eea
  where the bosons take an expectation value, 
and the fourier transform happens to give $\Phi^{(1)}_k$, the totally
symmetric Wannier state.  Magnetic frustration enables us to 
neglect the normally dominant interaction between the residual 
spin $\frac{1}{2}$ moments, however, we must still consider the 
Kondo coupling between localized and delocalized bands in the
expansion of the $Js_i\cdot s_j$ term of Eq.3: 
\begin{eqnarray}\label{}
H_{J} = J\sum_{<ij>} 
{\bf s}_{i}^{d}
\cdot 
{\bf s}_{j}^{c}
\longrightarrow
-\sum_{i,\Gamma}\frac{J^{\Gamma}}{2}(d_{i \alpha}^{\dagger}\psi_{i
\Gamma \alpha})(\psi^{\dagger}_{i \Gamma \beta}d_{i \beta}) 
\end{eqnarray}
where
the operators of the delocalized level have been expanded in terms of
orthogonal linear combinations as $\psi_{i \Gamma}^{\dagger} = \sum_j
\phi^{* \Gamma}(j-i) c^{\dagger}_j$ and $\phi^{* \Gamma}$ is a form
factor reflecting the phases at different nearest neighbor sites of
i, reflecting the possible arrangements of c$^{\dagger}$ on the
lattice.  This becomes: 
\begin{eqnarray}\label{}
H_{J}=
\sum_{i,\Gamma}\left[ 
(\Delta_{i \Gamma} \psi^{\dagger}_{i \Gamma
\beta}d_{i \beta} + \bar \Delta_{i \Gamma}d^{\dagger}_{i \beta}\psi_{i
\Gamma\beta }) + 2 \frac{ \bar \Delta_{i \Gamma}\Delta_{i
\Gamma}}{J^{\Gamma}}\right]
\end{eqnarray}
where we have performed a Hubbard-Stratonovich gauge transformation on the four-fermion term before fourier transforming to obtain the off-diagonal terms in Eq.(5).

 To treat the term of interest from a Hund's coupling:
\bea
H_{H} = \frac{J_H}{4}(c_{\alpha}^{\dagger}d_{\alpha}d_{\beta}^{\dagger}c_{\beta}) \rightarrow \frac{4}{J_H} \bar\xi \xi + i\bar\xi d_{\alpha}^{\dagger}c_{\alpha} + i c_{\alpha}^{\dagger}d_{\alpha} \xi
\eea
a Hubbard-Stratonovich gauge field has been introduced to decouple the four-fermion term{\cite{where}}.  Before the decoupling, $H_{H}|S=1> = 0|S=1>$ and $H_{H}|S=0> = \frac{J_H}{2}|S=0>$, so in the limit $J_H \rightarrow \infty$, the $|S=0>$ state will be forbidden. 
To write Eq.(4) we have taken the limit $J_H \rightarrow \infty$.  Although this appears to be non-Hermitian, the mean field values of $\xi$, $\bar\xi$ (independent variables) will also be imaginary so that physical quantities are real.








The Free energy can then be expressed as
\bea
F &=& -T\sum_{k,\sigma} 
Tr[ln(1 + e^{-\beta{\bf{h}}})] + 2\sum_{\Gamma}\frac{\bar
\Delta_{\Gamma}\Delta_{\Gamma}}{J^{\Gamma}} \cr
&+& \lambda_1(|b|^2 - 1) -  \lambda_2
\eea 
where ${\bf{h}}_k$ is the Hamiltonian expressed as a matrix as above.    Since $\bar \xi$ and $\bar \Delta$ both couple to ${\mathcal{C}}_{\alpha}^{\dagger}{\mathcal{D}}_{\alpha}$, changing one affects the other, such that $\frac{\partial^2 F}{\partial \bar\Delta \partial \Delta} \delta \bar\Delta +\frac{\partial^2 F}{\partial \bar\xi \partial \Delta}\delta\bar\xi = 0$. Integration over such fluctuations yields $\frac{1}{J^{*}} = \frac{1}{J} - \frac{F_{\bar \Delta \xi}F_{\bar \xi \Delta}}{2 F_{\xi \xi}}$, which guarantees the orthogonality of the second channel.  Channels with $F_{\bar\Delta \xi} \ne 0$ are effectively removed from consideration.  For example, on the square lattice at $\frac{1}{4}$-filling of the upper level, in the absence of the Hund's term, $\Phi^1$ would have the highest T$_{c2}$ while including the Hund's term the calculated T$_{c2}$ for this channel turns out to be negative.  That is no transition would occur if channels with F$_{\bar\Delta \xi} = 0$ did not exist.

To solve for T$_{c2}$, it is helpful to isolate the interaction contribution to the Green's functions for the delocalized level
\bea
G_c^{-1} = (G_c^0)^{-1} - \Sigma
\eea
where $\Sigma = \includegraphics[scale=0.5]{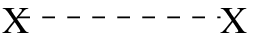} = |\Delta_{\Gamma}|^2\Phi_k^{\dagger \Gamma}(\omega - \lambda_2)^{-1}\Phi_k^{\Gamma}$ describes the interactions with the localized level.  Writing the free energy contribution as $F_{|\Delta_{\Gamma}|^2} = -T Tr\ln(G_c^{-1}) + \frac{2|\Delta_{\Gamma}|^2}{J^{\Gamma}}$ we find,
\bea
&\frac{\partial^2F}{\partial \Delta_{\Gamma}\partial\bar\Delta_{\Gamma}}|&_{ \Delta_{\Gamma}=i\xi=0} =  \frac{2}{J^{\Gamma}} + T \sum_{n,k} Tr[\Phi_k^{\dagger \Gamma}G_c^0G_d^0\Phi_k^{\Gamma}] \\ \nonumber &=& \sum_k \sum_{\alpha = 1}^{N} \frac{f(E_{k \alpha}) - f(\lambda_2)}{\lambda_2 - E_{k \alpha}} \eta_{k \alpha}^{\dagger}\Phi_k^{\dagger \Gamma}\Phi_k^{\Gamma}\eta_{k \alpha} + \frac{2}{J^{\Gamma}} = 0
\eea
where $E_{k \alpha}$ is the energy of the $\alpha$-th band of the free conduction electron problem and we have projected $\Phi_k^{\dagger}\Phi_k$ onto the eigenvectors of $H_{kin}$; $\mu,\lambda_1,\lambda_2$ are set by saddle-point evaluations to fix the filling.

\begin{figure}
\includegraphics[scale=0.7]{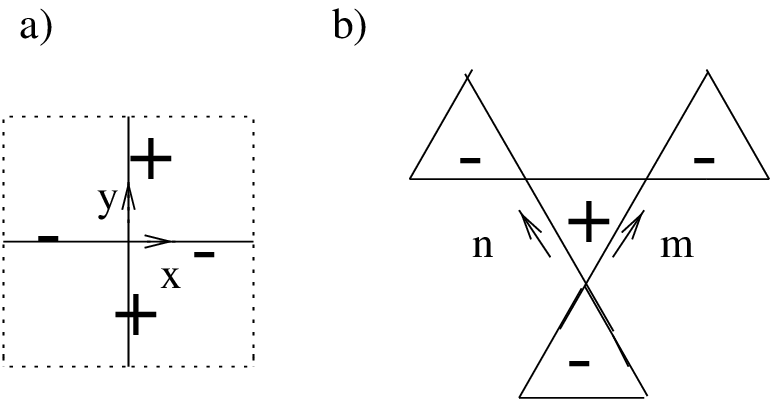}
\includegraphics[scale=0.7]{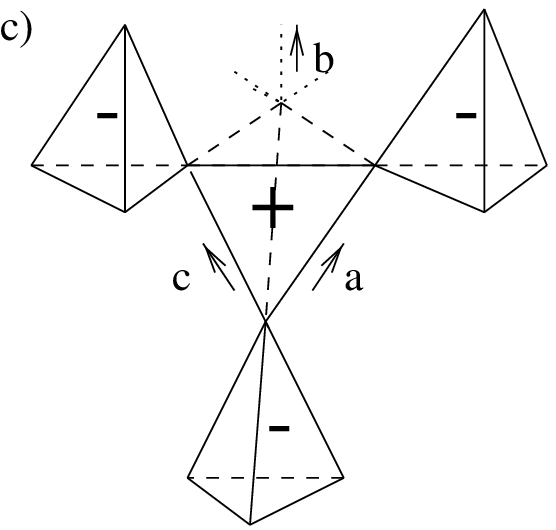} \caption{\label{figure} 
Symmetry  selects a unique symmetry channel for the second-stage Kondo effect
in the (a) square lattice (b) Kagom{\'e} lattice  and 
(c)Pyrochlore lattice. The $\pm$ signs refer to the relative phases of
the Wannier state at neigboring sites. 
}
\end{figure}

On a square lattice, C$_4$ symmetry means that the possible overlap phases with the nearest four neighbors are $\phi^{\lambda} = \frac{1}{2}(1, i^{\lambda}, i^{2\lambda}, i^{3\lambda})$, $\lambda = \{0,..,3\}$.  Hopping occurs in the first channel ($\Gamma = 1$) coinciding with the totally symmetric Wannier state $\lambda = 0$.   Time reversal symmetry specifies the d-wave like $\lambda = 2$ as the second channel ($\Gamma = 2$).  The form factors arise as:
$\sum_a \phi^{* 1}(a) c^{\dagger}_{i + a} =\frac{1}{2}( c^{\dagger}_{i + \hat x} + c^{\dagger}_{i - \hat y} + c^{\dagger}_{i - \hat x} + c^{\dagger}_{i + \hat y})$
which under fourier transform, defines $\Phi^{1}$:
$\psi_{i 1}^{\dagger} = \sum_k c^{\dagger}_k e^{-ikx_i} (cos(k_x) + cos(k_y)) = \sum_k c^{\dagger}_k e^{-ikx_i} \Phi^{* 1}_k$. Similarly, $\Phi^{* 2}_k = c(k_y) - c(k_x)$ are found  $\Phi^{* 4,3}_k = s(k_x) \pm is(k_y)$ completing the basis.

The effective T$_{c_2}$ is found from 
\bea
\sum_{k \sigma}\frac{f(0) - f(E_{qp})}{E_{qp}} \Phi^{* 2}_k\Phi^{2}_k + \frac{2}{J} = 0
\eea
where $E_{qp} = -2t|b|^2\Phi_k^{(1)} + \lambda_1 - \mu$ is determined self-consistently subject to the conditions $\lambda_2 = 0$ and $\sum_k n_f(-2t|b|^2\Phi_k^{(1)} + \lambda_1 - \mu) = \frac{1 - |b|^2}{2}$.




On the pyrochlore lattice, 4 bands and 6 nearest neighbors imply 24
independent symmetry channels!  The requirement that the states
preserve the equivalence between sites reduces this number to 6, given by
\[
\phi^{\pm}_{\alpha }=
\frac{1}{\sqrt{6}}(1,\alpha ,\alpha ^{2}, \pm 1,\pm \alpha ,\pm \alpha
^{2})
\]
where $\alpha$ is a cube root of unity.
Of these, only the state $\phi ^{-}= \frac{1}{\sqrt{6}}
(1,1,1,-1,-1,-1)$ is  orthogonal to the original channel 
and time-reversal invariant, uniquely specifying the
second channel.  This particular Wannier state has the property that
it changes sign between neigboring tetrahedra on the lattice, as illustrated
in Fig. 2. The four by four matrix $\Phi^{(1)}_k$ now has the following parts: 
\bea
\Phi^{(1)}_k =  \left(\begin{smallmatrix}0 & c(k_a) & c(k_b) & c(k_c) \cr c(k_a) & 0 & c(k_a - k_b) & c(k_a - k_c) \cr c(k_b) & c(k_a - k_b) & 0 & c(k_b - k_c) \cr c(k_c) & c(k_a - k_c) & c(k_b - k_c) & 0\end{smallmatrix}\right)
\eea
while, 
\bea
\Phi_k^{(2)} =  i\left(\begin{smallmatrix}0 & s(k_a) & s(k_b) & s(k_c) \cr -s(k_a) & 0 & s(k_a - k_b) & s(k_a-k_b) \cr -s(k_b) & -s(k_a - k_b) & 0 & s(k_b - k_c) \cr -s(k_c) & -s(k_a - k_c) & -s(k_b - k_c) & 0\end{smallmatrix}\right) 
\eea

For this case, the free hopping Hamiltonian has been treated
previously by Reimers et al.{\cite{berlinsky}} There are two flat
bands lying above two dispersing bands.  Strictly speaking, 
since we have $\frac{1}{2}$
an electron per site involved and four sites, this
would actually give us an insulator.  However,  since 
we have dropped the second degenerate band, a more reasonable band
structure would likely include band crossings and incomplete fillings
of the levels along the lines of Singh et al.{\cite{singh}}


A large body of experimental work has been done, and we now discuss a
simple interpretation of some of this work.  We have four localized
half-filled electron bands interacting via a Kondo-like hybridization
with four itinerant levels, each less than quarter-filled, which leaves us
with a hole-like Fermi surface below T$_{c2}$.  In this way, our model
is consistent with the 
observed sign change in the
Hall constant which has been observed at 50 K in
single crystal samples.{\cite{takagi}} A simple phenomenological fit
consistent with the ideas we have introduced for the resistivity
yields \bea \rho = \begin{cases}
	\rho_i + A_h T^2 			& T < 2K \\
	\rho_o + \gamma_{ph}T + A_p T^2		& T > 100K
\end{cases}
\eea
where $\rho_i = 21\mu\Omega cm$ (impurity scattering in the single crystal), $A_h = 2.0\frac{\mu\Omega cm}{K^2}$ (arises from the heavy Fermi liquid) as given in ref.4, and $\rho_o = 0.27m\Omega cm$ (impurity scattering from the localized spin $\frac{1}{2}$), $\gamma_{ph} = 1.1 \frac{\mu \Omega cm}{K}$ (phonon contribution above $\theta_{D}$) and $A_p = 2.7*10^{-3}\frac{\mu \Omega cm}{K^2}$ (arising from the 'porcupine' Fermi sea) have been extracted from Urano et al.'s data{\cite{takagi}}.  Since $\frac{A_h}{A_p} \approx (\frac{m^*_h}{m^*_p})^2$, this yields an effective mass ratio of $\frac{m^*_h}{m^*_p} \approx 27$.  An analogous fit of the magnetic susceptibility yields
\bea
\chi T = \begin{cases}
	\chi_h T				& T < 2K \\
	\chi_p (T-350K) + \frac{2 N_A(\mu g)^2 S(S+1) T}{k_B(T + \theta)} & T > 400 K
\end{cases}
\eea 
yields $\frac{\chi_h}{\chi_p} \approx \frac{m^*_h}{m^*_p} \approx
29$ where the $S = \frac{1}{2}$ ($\theta = 66K$, g = 2.21)
contribution has been first subtracted away from the high temperature
susceptibility measurements{\cite{kondo,fujiwara,takagi,muhtar}}
leaving a contribution linear in T.  The observed low temperature
metal insulator transition is hardly surprising under pressure as
broken symmetry Wannier states lie close in energy to the chosen
$\phi^2$ for the second channel.  Invariance of the heat capacity to
30T is indicative that the magnetic coupling J driving the spin
quenching is of much higher energy than the Kondo temperature
T$\approx$50K.

In conclusion, we have presented a physical picture within which one
can understand the physics of the frustrated heavy fermion
LiV$_2$O$_4$.  A simple t-J model is capable of describing both the
high temperature mixed valent phase, and the low temperature coupling
between a Fermi sea and spin liquid.  As no time
reversal symmetry breaking has been observed, we have been able to
uniquely identify the symmetry of the low temperature quasiparticles.  Our
picture gives a simple qualitative interpretation of the unusual
resistivity, the change in sign of the Hall resistance observed and
should be used to predict the result of de Haas-van Alphen experiments
yet to be performed. This work should serve as a starting point for a
DMFT or band structure calculation which might describe the detailed
energetics involved.  In particular it would be interesting to follow
the compression of the lattice and see how the second channel symmetry
depends on nearest neighbor distance.




We would like to acknowledge useful discussions with H. Takagi, K. LeHur and A. Rosch that led to this picture.  This work was supported by National Science Foundation grant DMR 9983156.


\end{document}